\documentclass[12pt, a4paper]{article}

\usepackage{graphicx}

\usepackage{stmaryrd}
\usepackage{overcite}
\usepackage{verbatim}
\usepackage{achemso}
\usepackage{verbatim}

\usepackage{multirow}
\usepackage{amsmath}
\usepackage{amsfonts}
\usepackage{amsgen}
\usepackage{longtable}
\usepackage{lscape}
\usepackage{threeparttable}
\usepackage{supertabular}
\usepackage{multicol}
\usepackage{float}
\usepackage{multirow}
\usepackage{setspace}
\usepackage{rotating}
\usepackage{amssymb}
\usepackage{layout}
\usepackage{wrapfig}
\usepackage{subfig}
\usepackage{enumerate}

\begin{document}

\begin{center}

{\bf\large{Dual Band Electrodes in Generator-Collector Mode: Simultaneous Measurement of Two Species}}

\hspace{2cm}

{\bf\large Edward O. Barnes$^{a}$, Grace E. M. Lewis$^{b}$, Sara E. C. Dale$^{b}$, Frank Marken$^{b}$ and Richard G. Compton$^{a}$*}

*Corresponding author\\
Email:~richard.compton@chem.ox.ac.uk\\
Fax:~+44~(0)~1865~275410; Tel:~+44~(0)~1865~275413.\\

$^{a}$Department~of~Chemistry, Physical~and~Theoretical~Chemistry~Laboratory, Oxford~University,
South~Parks~Road, Oxford, OX1~3QZ, United~Kingdom.\\
$^b$Department~of~Chemistry, University~of~Bath, Bath, BA2~7AY, United~Kingdom.
\vspace{1cm}

To be submitted as an article to:\\
\emph{The Journal of Electroanalytical Chemistry}

NOTICE: this is the author's version of a work that was accepted for publication in \emph{The Journal of Electroanalytical Chemistry}. Changes resulting from the publishing process, such as peer review, editing, corrections, structural formatting, and other quality control mechanisms may not be reflected in this document. Changes may have been made to this work since it was submitted for publication. A definitive version was subsequently published in \emph{The Journal of Electroanalytical Chemistry}, DOI: http://dx.doi.org/10.1016/j.jelechem.2013.05.022

\end{center}

\clearpage

\section*{Abstract}

A computational model for the simulation of a double band collector-generator experiment is applied to the situation where two electrochemical reactions occur concurrently. It is shown that chronoamperometric measurements can be used to take advantage of differences in diffusion coefficients to measure the concentrations of both electroactive species simultaneously, by measuring the time at which the collection efficiency reaches a specific value. The separation of the electrodes is shown to not affect the sensitivity of the method (in terms of percentage changes in the measured time to reach the specified collection efficiency), but wider gaps can provide a greater range of (larger) absolute values of this characteristic time. It is also shown that measuring the time taken to reach smaller collection efficiencies can allow for the detection of smaller amounts of whichever species diffuses faster. The case of a system containing both ascorbic acid and dopamine in water is used to exemplify the method, and it is shown that mole fractions of ascorbic acid between 0.055 and 0.96 can, in principle, be accurately measured.

\section*{Keywords}
Dual microbands; Generator-collector systems; Simultaneous concentration measurements; Potential step chronoamperometry.

\clearpage

\section{Introduction}

Generator-collector double electrode systems have found great utility in electrochemistry\cite{Barnes2012} since their conception in the late 1950s\cite{Frumkin1959}. In general, a generator-collector experiment consists of a reduction (or oxidation) at the generator electrode:
\begin{equation}
\mathrm{A} \pm \mathrm{e^{-}} \rightleftharpoons \mathrm{B}
\end{equation}
with a subsequent reaction, often the reverse of the first one, occurring at the collector electrode:
\begin{equation}
\mathrm{B} \mp \mathrm{e^{-}} \rightleftharpoons \mathrm{A}
\end{equation}
The collection efficiency, $N$, is defined as the ratio of the current measured at the collector electrode to that measured at the generator electrode:
\begin{equation}
N = -\frac{I_\mathrm{col}}{I_\mathrm{gen}}
\end{equation}
Measuring the collection efficiency can give information on the species involved in the reactions. If, for example, species B in the above scheme is unstable, and decomposes before it can reach the collector electrode, the collection efficiency will decrease, giving a measure of the rate at which the intermediate reaction occurs\cite{Albery1966d, Albery1966e}.

A great variety of geometries of generator-collector systems have been studied, including Nekresov's original rotating ring disc\cite{Frumkin1959} (also extensively studied by Albery and coworkers\cite{Albery1966, Albery1966a, Albery1966b, Albery1966c, Albery1966d, Albery1966e, Albery1966f, Albery1967, Albery1968, Albery1969, Albery1971a, Albery1971b, Albery1971c, Albery1971d,Albery1971e, Albery1972, Bruckenstein1977a, Albery1978, Albery1979, Albery1982, Albery1983b, Albery1989, Albery1989a, Albery1989b}), wall jet ring discs\cite{Brett1989, Brett1991, Compton1992}, dual band electrodes (used both in static solution\cite{Rajantie2001a, Rajantie2001, Rajantie2001b, Amatore2006} and hydrodynamic electrochemistry in a flow cell\cite{Aoki1977, Fisher1991, Compton1988c, Cooper1998}), dual discs\cite{Phillips1997, Baur2004, Cutress2011} and hemispheres\cite{Fulian1999, French2009b, French2009a, Dale2011}, and ring- or plane-recessed disc electrodes \cite{Menshykau2009, Menshykau2009a, Zhu2011}.

The uses of collector-generator systems have proved at least as varied as their geometries! In addition to the probing of follow up kinetics mentioned above, mechanistic details can be elucidated, for example the extent to which hydrogen peroxide plays a role as an intermediate in the reduction of oxygen to water\cite{Damjanovic1966, Damjanovic1967, Damjanovic1967a}. Other areas in which generator-collector systems have been successfully employed include detection of micromolar concentrations\cite{Dale2011}, studies of triple phase boundaries\cite{French2009b}, and measurement of one species in the presence of other, interfering, species.

Dual microband electrodes have previously been used by Williams \emph{et. al.}\cite{Rajantie2001a, Rajantie2001, Rajantie2001b} in electrochemical titration experiments, where a species which has been electrogenerated at the generator electrode undergoes a reaction with a target analyte in solution. Any surviving electrogenerated species is then detected at the collector electrode:
\begin{eqnarray}
\mathrm{A} + \mathrm{e^-} & \rightarrow & \mathrm{B} \quad \text{Generator electrode} \\
\mathrm{B} + \mathrm{W} & \rightarrow &  \mathrm{Z} \quad \text{Solution} \\
\mathrm{B} - \mathrm{e^-} & \rightarrow & \mathrm{A} \quad \text{Collector electrode}
\end{eqnarray}
Measuring the collection efficiency as a function of time yields information on the rate of the intermediate reaction, and hence the concentration of target analyte, W. Further studies by Amatore \emph{et. al.}\cite{Amatore2006} have theoretically investigated applying potential pulses to the generator electrode, and using the response at the collector electrode to obtain a time of flight for the electrogenerated species, and hence a measure of its diffusion coefficient.

\begin{figure}[h]
\begin{center}
\includegraphics[width = 0.9\textwidth]{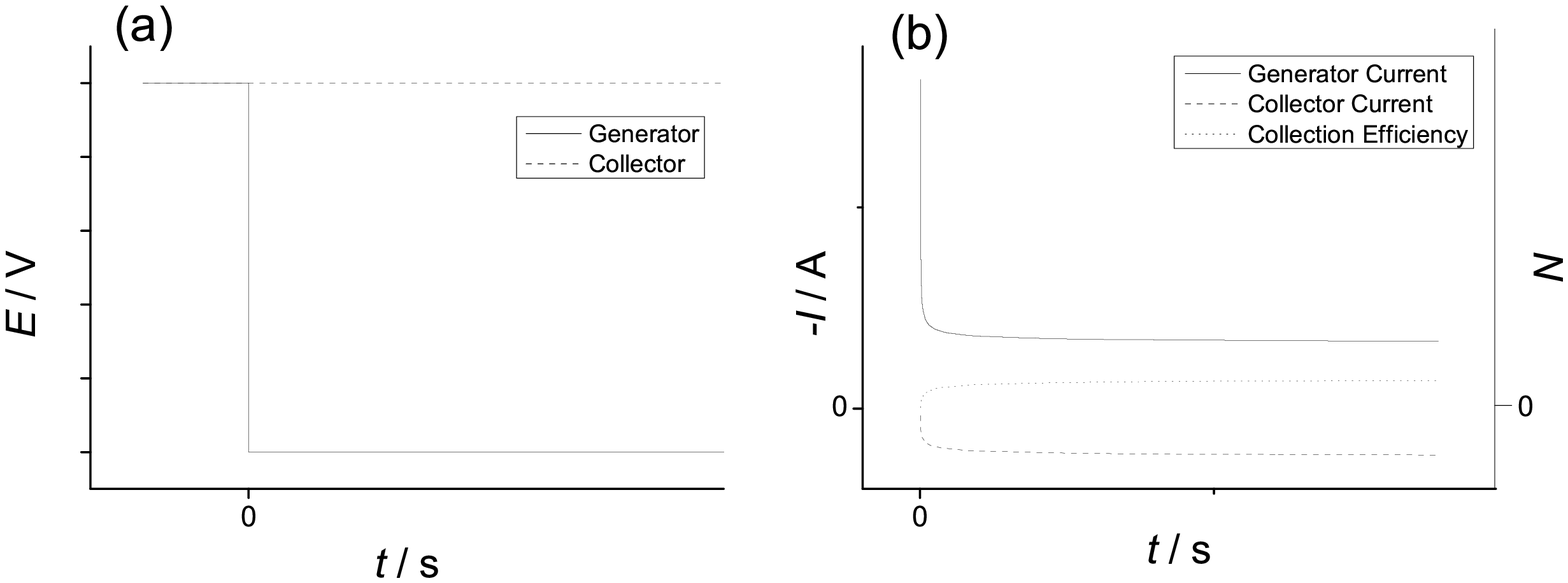}
\caption{Schematic diagram of (a): The potentials applied to the generator and collector electrodes during a chronoamperometric experiment and (b): The current responses and the collection efficiency.} \label{EXPERIMENT SCHEMATIC}
\end{center}
\end{figure}

In this paper, we consider a chronoamperometric experiment using \emph{dual} microband electrodes, with two electroactive species initially present in solution, A and X. Before the experiment starts, the potentials applied to both electrodes are such that no reduction occurs. At the beginning of the experiment, the potential applied to the generator electrode is stepped to a value such that reduction of the electroactive species A and X occurs at a mass transport controlled rate, forming B and Y. The potential at the collector electrode, however, remains at a valu to re-oxidise the electrogenerated species at a mass transport controlled rate. Schematic representations of the applied potentials, the currents produced, and the collection efficiency as the experiment progresses are shown in Figure \ref{EXPERIMENT SCHEMATIC}. The collection efficiency is seen to increase with time, and continues to do so until it approaches 1 as time $t \to \infty$. The rate at which it increases will be dependent on how quickly the intermediate species, B and Y, diffuse across the inter-electrode gap. If all species diffuse at the same rate, then the relative initial concentrations of A and X will have no effect on how the collection efficiency varies with time.

If, however, the diffusion coefficients of the species produced at the generator are different, they will traverse the gap between the two electrodes at different rates, and thus the collection efficiencies attributable to each reaction will increase at different rates. The overall measured collection efficiency will therefore depend upon the relative concentrations of the two species. This study investigates the extent to which this is the case, and explores the application of this to the simultaneous determination of the concentration of both electroactive species in solution.

\section{Theory}

\begin{figure}[h]
\begin{center}
\includegraphics[width = 0.9\textwidth]{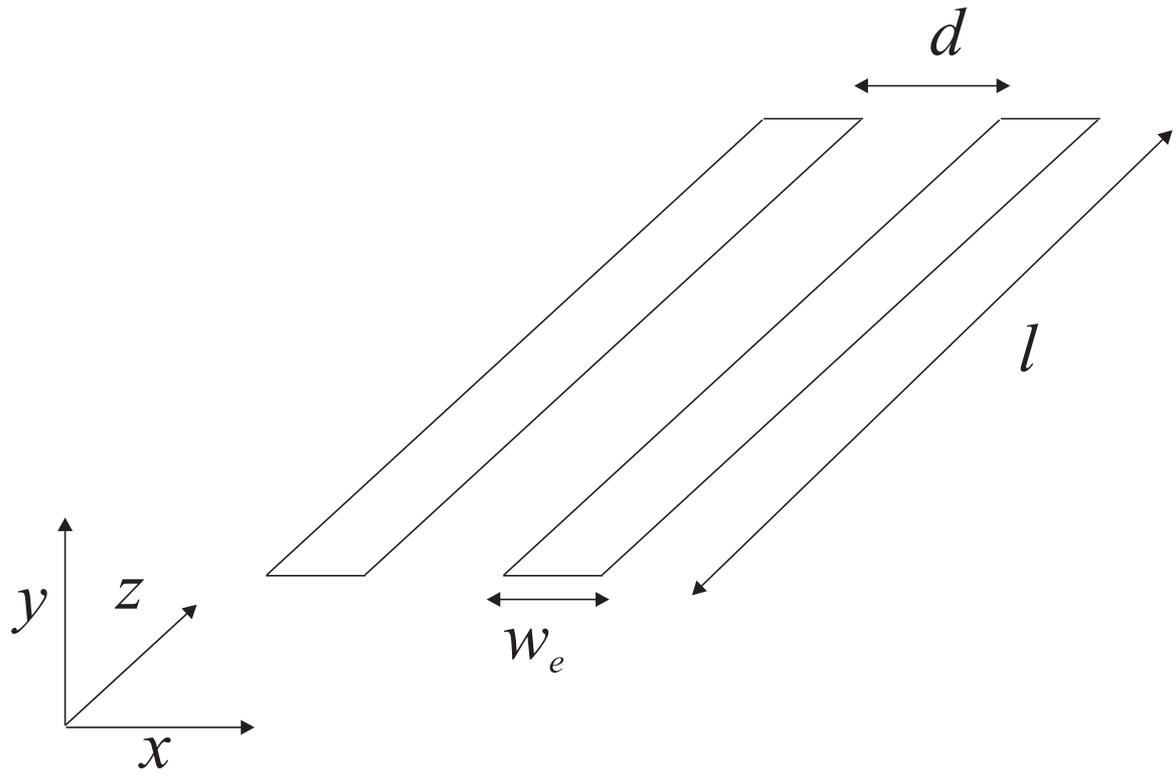}
\caption{Schematic diagram of a dual microband electrode system} \label{DUAL BAND SCHEMATIC}
\end{center}
\end{figure}

In this paper we consider potential step chronoamperometry at dual microband electrodes. Two concurrent one electron reductions are considered to occur:
\begin{eqnarray}
\mathrm{A} + \mathrm{e}^- \rightleftharpoons \mathrm{B} \label{REDUCTION1}\\
\mathrm{X} + \mathrm{e}^- \rightleftharpoons \mathrm{Y} \label{REDUCTION2}
\end{eqnarray}
The dual microband electrodes are considered to be parallel and effectively of infinite length. One electrode (the generator electrode) has an applied potential such that the reductions both occur at a mass transport controlled rate, and the other electrode (the collector electrode) has an applied potential such that the reverse, oxidation, reactions occur at a mass transport controlled rate. A schematic diagram of the dual microband electrodes is shown in Figure \ref{DUAL BAND SCHEMATIC}, which also defines the Cartesian coordinates.

By considering the case where the length of the electrodes, $l$, is much greater than their width, $w_e$, or the distance between them, $d$, the mass transport equation becomes two dimensional. Assuming an excess amount of inert supporting electrolyte is present, such that mass transport is diffusion only, the mass transport equation is:
\begin{equation}
\frac{\partial{c_\mathrm{i}}}{\partial{t}} = D_\mathrm{i}\left(\frac{\partial^2c_\mathrm{i}}{\partial{x^2}} + \frac{\partial^2c_\mathrm{i}}{\partial{y^2}}\right)
\end{equation}
where $c_\mathrm{i}$ is the concentration of species i (mol m$^{-3}$), $D_\mathrm{i}$ is the diffusion coefficient of species i (m$^2$ s$^{-1}$), and $x$ and $y$ are the coordinates defined in Figure \ref{DUAL BAND SCHEMATIC} (m). All symbols are defined in Table \ref{DIMENSIONAL}.

\subsection{Boundary conditions}

\begin{figure}[h]
\begin{center}
\includegraphics[width = 0.9\textwidth]{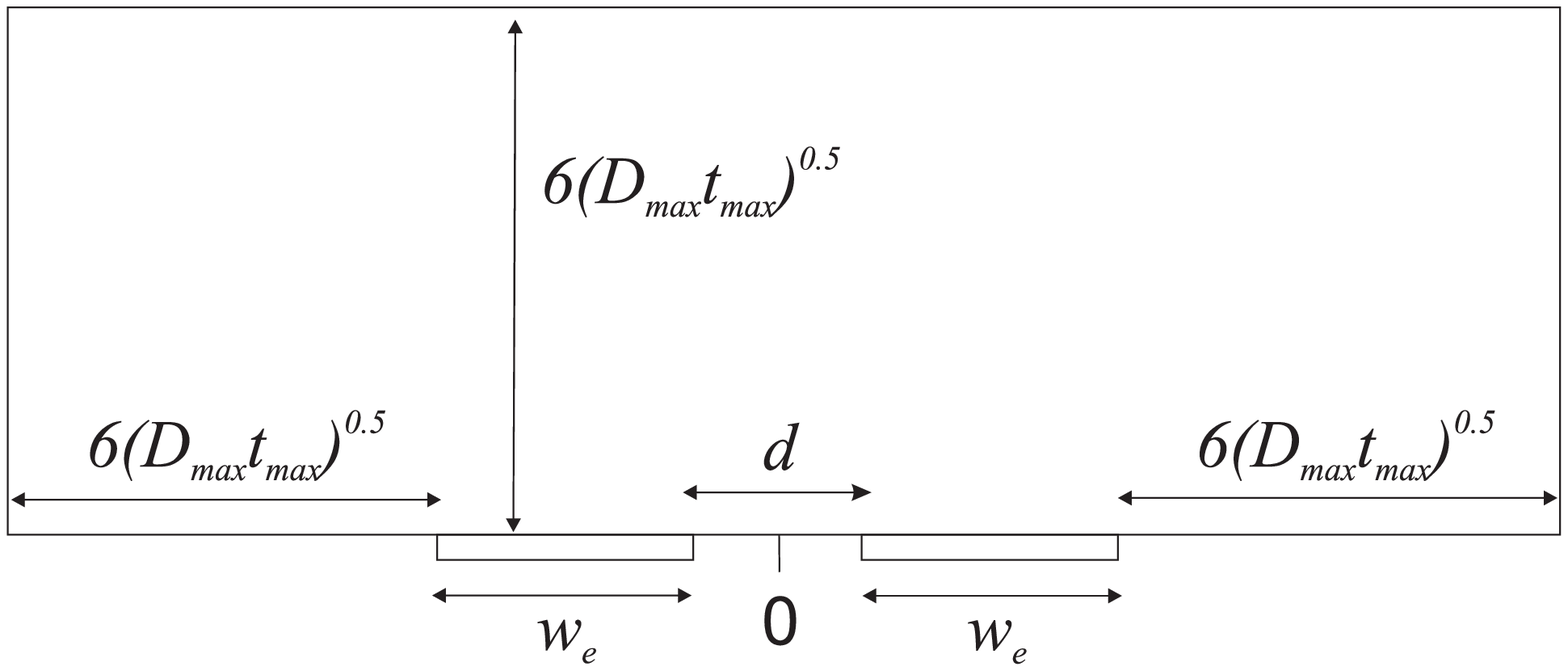}
\caption{Schematic diagram of the simulation space used in this study} \label{SIMULATION SPACE}
\end{center}
\end{figure}

Initially, throughout solution, species A and X are present in their bulk concentrations, while species B and Y are not preset at all:
\begin{equation}
t < 0\text{; all }x\text{; all }y\text{;} \quad c_\mathrm{A} = c_\mathrm{A}^{*} \quad c_\mathrm{B} = 0\quad c_\mathrm{X} = c_\mathrm{X}^{*} \quad c_\mathrm{Y} = 0
\end{equation}
At time $t<0$, the potentials at both electrodes are such that no reactions occur, neither species A nor X are reduced. At $t=0$ the generator electrode potential is stepped to a value such that species A and X are reduced at a mass transport controlled rate. The collector electrode potential remains at the same value, such that species B and Y are re-oxidised at a mass transport controlled rate. The generator electrode boundary condition is therefore:
\begin{equation}
t \geq 0\text{; }-w_e-\frac{d}{2} \leq x \leq -\frac{d}{2}\text{; }y = 0
\begin{cases}
c_\mathrm{A} = 0\\
D_\mathrm{B}\left(\frac{\partial{c_\mathrm{B}}}{\partial{y}}\right) = -D_\mathrm{A}\left(\frac{\partial{c_\mathrm{A}}}{\partial{y}}\right)\\
c_\mathrm{X} = 0\\
D_\mathrm{Y}\left(\frac{\partial{c_\mathrm{Y}}}{\partial{y}}\right) = -D_\mathrm{X}\left(\frac{\partial{c_\mathrm{X}}}{\partial{y}}\right)
\end{cases}
\end{equation}
and the collector electrode condition:
\begin{equation}
t \geq 0\text{; }\frac{d}{2} \leq x \leq w_e+\frac{d}{2}\text{; }y = 0
\begin{cases}
D_\mathrm{A}\left(\frac{\partial{c_\mathrm{A}}}{\partial{y}}\right) = -D_\mathrm{B}\left(\frac{\partial{c_\mathrm{B}}}{\partial{y}}\right)\\
c_\mathrm{B} = 0\\
D_\mathrm{X}\left(\frac{\partial{c_\mathrm{X}}}{\partial{y}}\right) = -D_\mathrm{Y}\left(\frac{\partial{c_\mathrm{Y}}}{\partial{y}}\right)\\
c_\mathrm{Y} = 0
\end{cases}
\end{equation}
On the insulating surface around (and between) the electrodes, a zero flux condition is imposed:
\begin{equation}
t \geq 0\text{; }y = 0\text{;}
\begin{cases}
-w_e-\frac{d}{2} < x\\
-\frac{d}{2} < x < \frac{d}{2} \quad\quad \left(\frac{\partial{c_\mathrm{i}}}{\partial{y}}\right) = 0\\
x > \frac{d}{2} + w_e
\end{cases}
\end{equation}
The bulk solution boundary is defined as being $6\sqrt{D_\mathrm{max}t_\mathrm{max}}$ from the electrode surfaces in the $x$ and $y$ directions, as shown schematically in Figure \ref{SIMULATION SPACE}, where $D_\mathrm{max}$ is the largest diffusion coefficient in the system and $t_\mathrm{max}$ is the duration of the experiment. This is known to be well outside the diffusion layer\cite{Gavaghan1998, Gavaghan1998a, Gavaghan1998b}. At these boundaries, concentrations are set at their bulk values. Defining $x_\mathrm{max}$ as $\frac{d}{2} + w_e + 6\sqrt{D_\mathrm{max}t_\mathrm{max}}$ and $y_\mathrm{max}$ as $6\sqrt{D_\mathrm{max}t_\mathrm{max}}$:
\begin{equation}
t \geq 0\text{; }
\begin{cases}
x = -x_\mathrm{max}\text{; all }y\\
x = x_\mathrm{max}\text{; all }y \quad\quad c_\mathrm{A} = c_\mathrm{A}^{*} \quad c_\mathrm{B} = 0\quad c_\mathrm{X} = c_\mathrm{X}^{*} \quad c_\mathrm{Y} = 0\\
\text{all }x\text{; }y = y_\mathrm{max}
\end{cases}
\end{equation}

\subsection{Normalised parameters}

The model can be simplified by introducing a series of normalised parameters, which remove scaling factors such as electrode size and bulk concentrations. A full list is given in Table \ref{DIMENSIONLESS}. After dimensional parameters have been substituted for normalised ones, the mass transport equation becomes:
\begin{equation}
\frac{\partial{C_\mathrm{i}}}{\partial{t}} = D'_\mathrm{i}\left(\frac{\partial^2C_\mathrm{i}}{\partial{X^2}} + \frac{\partial^2C_\mathrm{i}}{\partial{Y^2}}\right)
\end{equation}
The initial state of the system becomes:
\begin{equation}
\tau < 0\text{; all }X\text{; all }Y\text{;} \quad C_\mathrm{A} = 1 \quad C_\mathrm{B} = 0\quad C_\mathrm{X} = C_\mathrm{X}^{*} \quad C_\mathrm{Y} = 0
\end{equation}
After the potential step, the generator electrode boundary condition is:
\begin{equation}
\tau \geq 0\text{; }-1-\frac{d'}{2} \leq X \leq -\frac{d'}{2}\text{; }Y = 0
\begin{cases}
C_\mathrm{A} = 0\\
D'_\mathrm{B}\left(\frac{\partial{C_\mathrm{B}}}{\partial{Y}}\right) = -\left(\frac{\partial{C_\mathrm{A}}}{\partial{Y}}\right)\\
C_\mathrm{X} = 0\\
D'_\mathrm{Y}\left(\frac{\partial{C_\mathrm{Y}}}{\partial{Y}}\right) = -D'_\mathrm{X}\left(\frac{\partial{C_\mathrm{X}}}{\partial{Y}}\right)
\end{cases}
\end{equation}
and on the collector electrode:
\begin{equation}
\tau \geq 0\text{; }\frac{d'}{2} \leq X \leq 1+\frac{d'}{2}\text{; }Y = 0
\begin{cases}
\left(\frac{\partial{C_\mathrm{A}}}{\partial{Y}}\right) = -D'_\mathrm{B}\left(\frac{\partial{C_\mathrm{B}}}{\partial{Y}}\right)\\
C_\mathrm{B} = 0\\
D'_\mathrm{X}\left(\frac{\partial{C_\mathrm{X}}}{\partial{Y}}\right) = -D'_\mathrm{Y}\left(\frac{\partial{C_\mathrm{Y}}}{\partial{Y}}\right)\\
C_\mathrm{Y} = 0
\end{cases}
\end{equation}
The insulating surface boundary condition becomes:
\begin{equation}
\tau \geq 0\text{; }Y = 0\text{;}
\begin{cases}
-1-\frac{d'}{2} < X\\
-\frac{d'}{2} < X < \frac{d'}{2} \quad\quad \left(\frac{\partial{C_\mathrm{i}}}{\partial{Y}}\right) = 0\\
X > \frac{d'}{2} + 1
\end{cases}
\end{equation}
The bulk solution boundary is now defined to be $6\sqrt{D'_\mathrm{max}\tau_\mathrm{max}}$ from the electrodes, giving $X_\mathrm{max} = \frac{d'}{2} + 1 + 6\sqrt{D'_\mathrm{max}\tau_\mathrm{max}}$ and $Y_\mathrm{max} = 6\sqrt{D'_\mathrm{max}\tau_\mathrm{max}}$. Bulk boundary conditions therefore become:
\begin{equation}
\tau \geq 0\text{; }
\begin{cases}
X = -X_\mathrm{max}\text{; all }Y\\
X = X_\mathrm{max}\text{; all }Y \quad\quad C_\mathrm{A} = 1 \quad C_\mathrm{B} = 0 \quad C_\mathrm{X} = C_\mathrm{X}^{*} \quad C_\mathrm{Y} = 0\\
\text{all }X\text{; }Y = Y_\mathrm{max}
\end{cases}
\end{equation}

\subsection{Numerical methods}

To obtain a numerical solution to the problem outlined above, the mass transport equation and the boundary conditions are discretised according to the Crank Nicolson method\cite{Crank1947} and solved using the alternating direction implicit (ADI) method in conjunction with the Thomas algorithm\cite{Press2007}.

In addition to discretising the equations, spatial and temporal grids must be defined to solve the equations over. The temporal grid begins at $\tau_0 = 0$ and is initially linear, with a step size of $\Delta\tau$, until some switching time $\tau_s$ after which the grid expands:
\begin{eqnarray}
\tau < \tau_s \quad \tau_k & = & \tau_{k-1} + \Delta\tau \\
\tau \geq \tau_s \quad \tau_k & = & \tau_{k-1} + \gamma_\tau\left(\tau_{k-1} - \tau_{k-2}\right)
\end{eqnarray}
The spatial grid in the $Y$ direction simply expands away from the electrode/insulating surface. The first point, $Y_0$, is set as 0, and the second point some value $\Delta Y$. The rest of the grid is defined:
\begin{equation}
Y_j = \gamma_YY_{j-1}
\end{equation}
In the $X$ direction, the grid is similarly defined, and expands away from points located at both edges of each electrode, until it reaches either the centre of an electrode ($X=\pm\left(0.5 + \frac{d'}{2}\right)$), the centre of the inter-electrode gap ($X = 0$), or the edge of the simulation space. To achieve convergence, the following parameters were used: $\Delta\tau = 1\times 10^{-7}$, $\tau_s = 1\times 10^{-4}$, $\gamma_\tau = 1.0001$, $\Delta Y = 8 \times 10^{-5}$, $\gamma_Y = 1.125$. The model was programmed in C++ and all simulations carried out on an Intel(R) Xenon(R) 2.26 GHz PC with 2.25 GB RAM.

\section{Theoretical Results}

\subsection{Varying relative concentrations of species A and X}

If, during a double electrode potential step chronoamperometry experiment as described above, only one of reactions \ref{REDUCTION1} and \ref{REDUCTION2} above is considered to occur in isolation at the generator electrode, with the reverse reaction occurring at the collector electrode (both at mass transport controlled rates), then the rate at which the collection efficiency increases will be a function of the inter electrode distance, and the diffusion coefficients of the species involved. Species which diffuse more slowly will traverse the gap more slowly, resulting in a lower collection efficiency at any given time from the start of the experiment. So if two separate experiments are run, one with reaction \ref{REDUCTION1} and one with reaction \ref{REDUCTION2}, and species B and Y have different diffusion coefficients, then at any given time after the beginning of the experiment the measured collection efficiencies will be different. If both reactions occur together in the same experiment, the measured collection efficiency will be somewhere between the two individual cases.

\begin{figure}[h]
\begin{center}
\includegraphics[width = 0.9\textwidth]{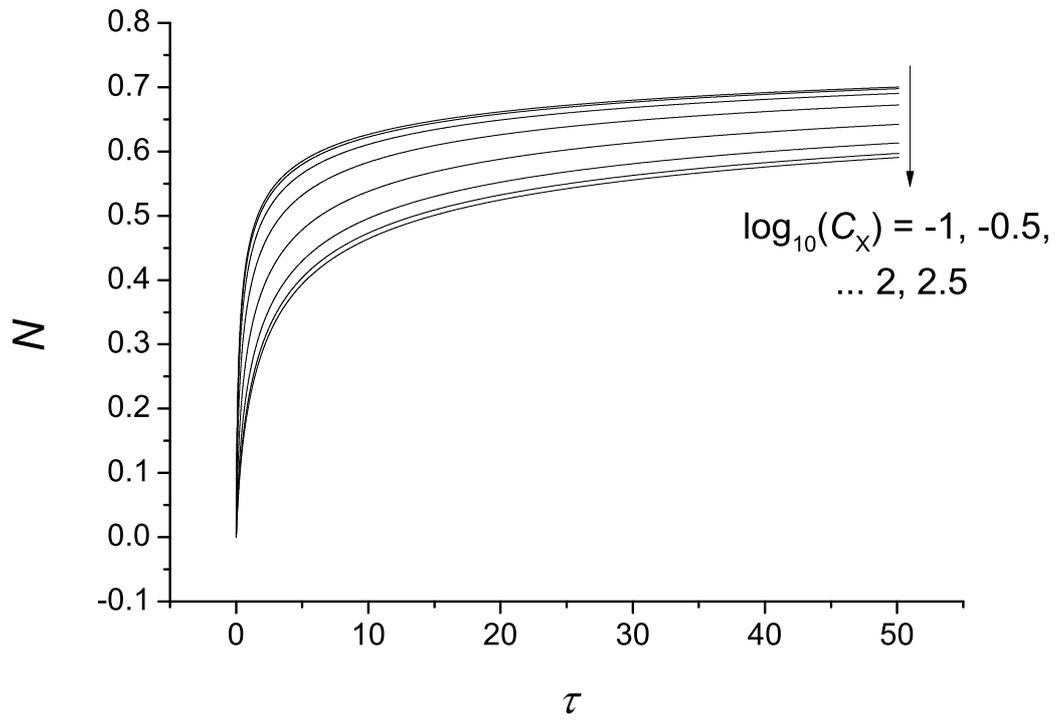}
\caption{Collection efficiencies for various values of $C_\mathrm{X}$ as a function of dimensionless time. $d' = 0.1$, $D'_\mathrm{B} = D'_\mathrm{A} \left(=1\right)$, $D'_\mathrm{Y} = D'_\mathrm{X} = 0.1$} \label{COLLECTION EFFICIENCIES}
\end{center}
\end{figure}

This is exemplified in Figure \ref{COLLECTION EFFICIENCIES}, which shows simulated collection efficiencies for a variety of relative initial concentrations of species A and B. In the simulations, $D'_\mathrm{A} = D'_\mathrm{B} = 1$, $D'_\mathrm{X} = D'_\mathrm{Y} = 0.1$ and $d' = 0.1$. It is seen that as $C_\mathrm{X}$ approaches zero, the measured collection efficiency curves begin to converge on each other, eventually reaching the result obtained when only species A is present in solution. Conversely, as $C_\mathrm{X}$ becomes very large, the results converge on the limit where only species X is initially present in solution.

Using this result, the relative concentrations of species A and X in a solution of unknown composition can be determined. If the time taken for the collection efficiency to reach some specified value, say 0.5, is measured, and compared to a calibration curve, the relative concentrations of species A and X can be estimated. The absolute, un-normalised values can then be calulated from the magnitudes of the collector and generator currents.

\begin{figure}[h]
\begin{center}
\includegraphics[width = 0.9\textwidth]{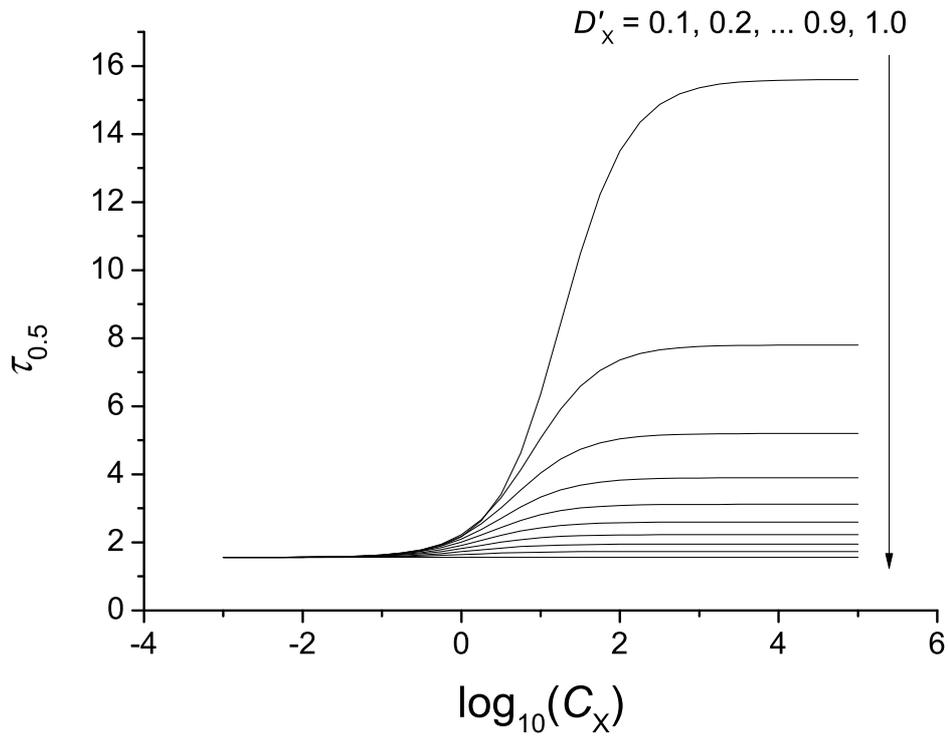}
\caption{$\tau_{0.5}$, the time taken for the collection efficiency to reach 0.5, as a function of log$_{10}(C_\mathrm{X})$, for various values of $D'_\mathrm{X}$. $d' = 0.1$, $D'_\mathrm{B} = D'_\mathrm{A} \left(=1\right)$, $D'_\mathrm{Y} = D'_\mathrm{X}$} \label{CALIBRATION CURVE}
\end{center}
\end{figure}

Figure \ref{CALIBRATION CURVE} shows the dimensionless time taken for the collection efficiency to reach 0.5, $\tau_{0.5}$, for various values of $C_\mathrm{X}$ and $D'_\mathrm{X}$. In each case, $d' = 0.1$, $D'_\mathrm{B} = D'_\mathrm{A}$ and $D'_\mathrm{Y} = D'_\mathrm{X}$. It is seen that the sensitivity of this method increases as the relevant diffusion coefficients become more and more different from each other (uppermost line in Figure \ref{CALIBRATION CURVE}). Conversely, when the diffusion coefficients are equal to each other, this method becomes completely insensitive to the solution composition. More specifically, it is seen that:
\begin{equation}
\frac{\tau_{0.5}^{C_{\mathrm{X}=0}}}{\tau_{0.5}^{C_{\mathrm{X}\to\infty}}} = \frac{D'_\mathrm{A}}{D'_\mathrm{X}}
\end{equation}
so if species A and B diffuse 10 times faster than species X and Y, $\tau_{0.5}$ increases 10 fold between the limits of only A initially present and only X initially present.

\subsection{Effect of Inter-Electrode Distance}

\begin{figure}[h]
\begin{center}
\includegraphics[width = 0.9\textwidth]{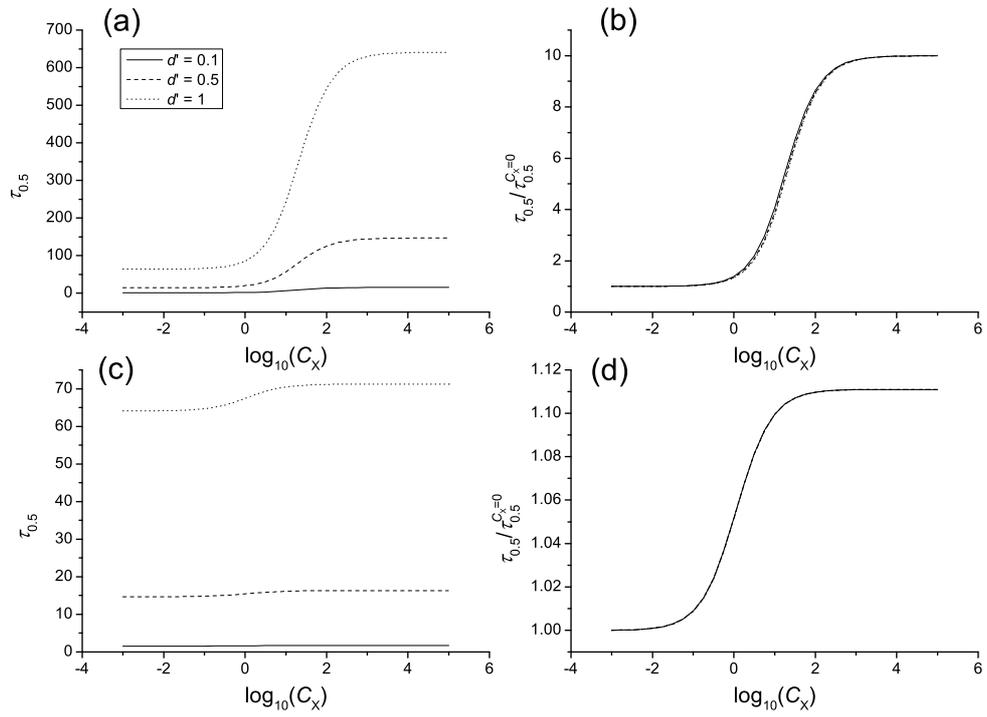}
\caption{(a): Absolute values of $\tau_{0.5}$ as a function of log$_{10}(C_\mathrm{X})$ for various gap sizes, with $D'_\mathrm{X} = 0.1$. (b): as (a), but $\tau_{0.5}$ values taken relative to its value when $C_\mathrm{X}=0$ for each gap size. (c) and (d) as (a) and (b), except with $D'_\mathrm{X} = 0.9$.} \label{VARY d'}
\end{center}
\end{figure}

By changing the value of $d'$, the effect of the inter-electrode distance on the sensitivity of this method to determine two concentration simultaneously can be investigated. Figure \ref{VARY d'} (a) shows the calibration curves obtained when $D'_\mathrm{X} = 0.1$ for the three dimensionless inter-electrode distances $d'=0.1$, 0.5 and 1. Figure \ref{VARY d'} (b) shows these curves normalised to the values of $\tau_{0.5}$ obtained when $C_\mathrm{X} = 0$, $\tau_{0.5}^{C_\mathrm{X}=0}$. (c) and (d) show the same, but now for $D'_\mathrm{X}=0.9$. It is seen from (a) and (c) that larger gaps between the electrodes produce larger absolute differences between the extremes of the calibration curves. The normalised curves, however, show that the values of $\tau_{0.5}$ taken relative to the lower extremes are completely insensitive to the inter-electrode distance, whether the diffusion coefficients are very different or nearly the same. This insensitivity (in the relative values) makes the ideal gap size a matter of experimental convenience. This may favour larger gap sizes which produce greater changes in the absolute values of $\tau_{0.5}$ across the calibration curve, although this must be balanced with the longer experimental timescales necessary with larger gaps.

\subsection{Effect of the Chosen Collection Efficiency}

\begin{figure}[h]
\begin{center}
\includegraphics[width = 0.9\textwidth]{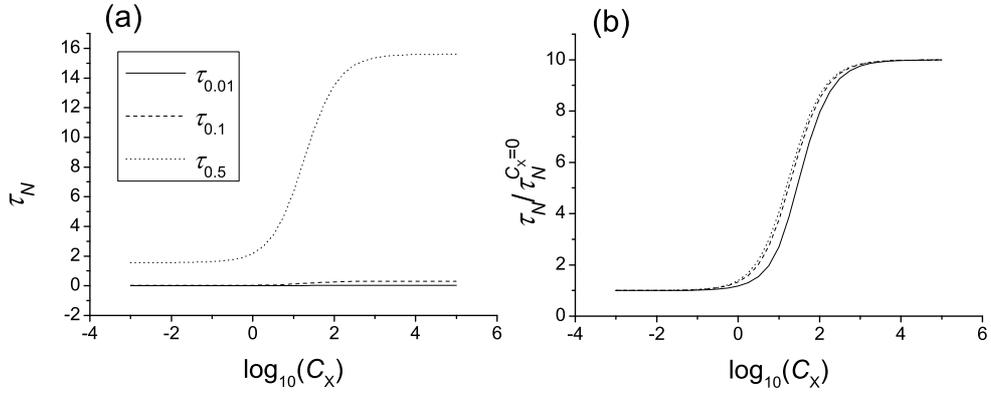}
\caption{(a): Absolute values of $\tau_{N}$ as a function of log$_{10}(C_\mathrm{X})$ for various values of $N$, with $d' = 0.1$ and $D'_\mathrm{X} = 0.1$. (b): as (a), but $\tau_{N}$ values taken relative to its value when $C_\mathrm{X}=0$ for each value of $N$.} \label{VARY TAUN}
\end{center}
\end{figure}

In the above considerations, a collection efficiency of 0.5 was arbitrarily chosen as the point at which to construct the calibartion curves. Other points may however be used. Figure \ref{VARY TAUN} shows the calibration curves obtained by taking measurements at collection efficiencies of 0.01, 0.1 and 0.5, corresponding to $\tau_N=\tau_{0.01}$, $\tau_{0.1}$ and $\tau_{0.5}$ respectively. Figure \ref{VARY TAUN} (a) shows the absolute $\tau_N$ values taken at each point, and Fiure \ref{VARY TAUN} (b) shows these values relative to the lower extremes where $C_\mathrm{X} = 0$, $\tau_N^{C_\mathrm{X}=0}$. In each case, $D'_\mathrm{A} = D'_\mathrm{B} = 1$, $D'_\mathrm{X} = D'_\mathrm{Y} = 0.1$ and $d'=0.1$. The un-normalied plots show that calibration curves taken at higher collection efficiencies produce a much larger range of absolute $\tau$ values, with smaller collection efficiencies producing extremely small values of $\tau$ which may prove difficult to measure experimentally ($\tau_{0.01}$ varies between the extremes of 0.003689 and 0.03689). However, the normalised plots in Figure \ref{VARY TAUN} (b) show that using a smaller collection efficieny, here 1\%, can make the calibration curve sensitive to smaller (relative) concentrations of species A, the faster diffusing species (\emph{i.e.} higher values of $C_\mathrm{X}$). If, therefore, the difficulties in accurately measuring such small times can be overcome, using a small collection efficiency as the basis of the calibration curve may be desirable if one of the two species (the faster diffusing one) is present in trace quantities.

\subsection{Theoretical Application to Simultaneous Analysis of Ascorbic Acid and Dopamine}

\begin{figure}[h]
\begin{center}
\includegraphics[width = 0.9\textwidth]{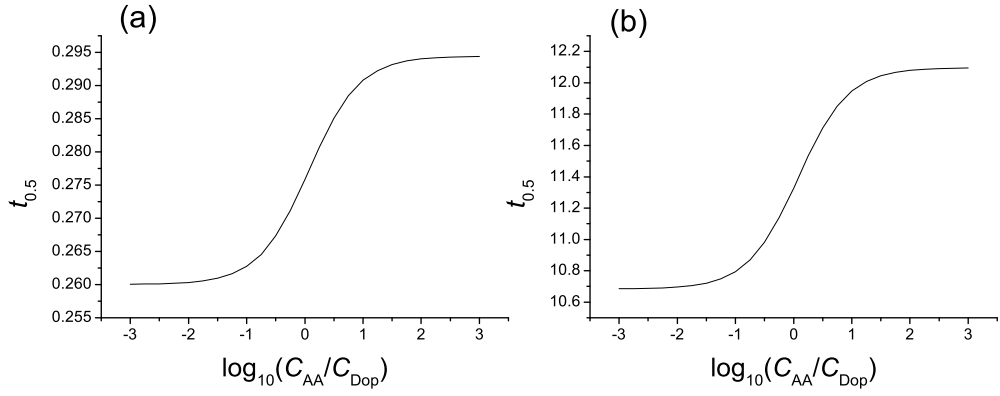}
\caption{$t_\mathrm{0.5}$ as a function of log$_{10}(\frac{C_\mathrm{AA}}{C_\mathrm{Dop}})$. Electrode widths, $w_e = 10$ $\mu$m. (a): Gap size, $d = 1 \mu$m, (b): Gap size, $d = 10$ $\mu$m.} \label{AA AND DOP}
\end{center}
\end{figure}

Ascorbic acid (AA) and dopamine (Dop) are biologically important molecules, both vital to human health,  and a great deal of research has been conducted to detect one in the presence of the other\cite{Zhu2011, Oliveira2013, Zhang2013}. To show how this method of measuring the concentration of two species simultaneously can be used, these two molecules are taken as examples. Dual band chronoamperometry is simulated, oxidising AA and Dop at the generator electrode and re-reducing them at the collector. The diffusion coefficients of AA and Dop in water have been measured as $0.53 \times 10^{-5}$ and $0.60 \times 10^{-5}$ cm$^{2}$ s$^{-1}$ respectively\cite{Gerhardt1982}, and the diffusion coefficients of the reduced species are assumed to be the same as for the parent species. The band electrodes are taken to be 10 $\mu$m in width, and two gap sizes are considered, 1 $\mu$m and 10 $\mu$m. The simulated calibration curves produced are shown in Figure \ref{AA AND DOP}. For both inter-electrode distances, the central 90\% of the $\tau_{0.5}$ range covers values of $c_\mathrm{AA}$/$c_\mathrm{Dop}$ ranges of 0.058 to 21.88 (corresponding to mole fractions of AA of 0.055 to 0.96).

\section{Summary of Method}

The steps involved in using this method to simultaneously determine two concentrations can be summarised as follows:
\begin{enumerate}[i)]
\item
Construct\cite{Williams1997} or purchase dual microband electrodes of known dimensions. A larger inter-electrode distance will result in a greater range of times to reach a given collection efficiency, although the relative sensitivity is not affected by the inter-electrode distance.
\item
If the diffusion coefficients of species A, B, X and Y are unknown, these should be measured using single and/or double potential step chronoamperometry at a single electrode (\emph{e.g.} a microdisc).
\item
Construct a calibration curve of relative concentration plotted against the time taken for the collection efficiency to reach a certain value when performing potential step chronoamperometry at dual microband electrodes. This can be done computationally as outlined above, or experimentally using solutions of known relative concentrations. Choosing a smaller collection efficiency will allow for more accurate measurement of trace amounts of the faster diffusing species, but will also require smaller timescales to be accurately measured.
\item
The time taken for the collection efficiency to reach the specified value can then be measured for a solution of unknown composition, and the relative concentrations determined from the calibration curve.
\end{enumerate}

It should be noted that in this work, only analytes which do not undergo follow up homogeneous kinetics are considered. The effect of homogeneous kinetics would lower the collection efficiencies, and will add some difficulties in selecting an appropriate collection efficiency target and electrode geometry (the previously advantageous larger gap size would now only further decrease collection efficiencies as follow up kinetics have longer to progress). Further, the scope of this work only includes electron transfers in the infinitely fast Nernstian limit. The effect of slower electron transfer on chronoamperometry has been studied by this group previously\cite{Belding2009}, and its application to this method is facile.

\section{Conclusions}

In this study, we have demonstrated the use of dual microband electrodes in chronoamperometric generator/collector mode to detect the concentrations of two competing electroactive species simultaneously, provided the diffusion coefficients involved are known or separately measured, and are not equal to each other. An example case of ascorbic acid and dopamine in water was used, and a theoretical range of detectable mole fractions of ascorbic acid was found to be between 0.055 and 0.96. It has been shown that the timescales involved using dual bands in this method are reasonable and easily measurable, whereas an interdigitated array, for example, would produce high collection efficiencies at much shorter and harder to measure times. A further possible example where this could be usefully applied would be for detecting the relative concentrations of two enantiomers, which could be given different diffusion coefficients by the use of a chiral solvent. After calibration, the experimental ease of this method makes it ideal for use in detecting the individual concentrations of two otherwise difficult to discriminate species in a quick and cheap manner.

\section*{Acknowledgments}

EOB and SECD thank EPSRC, and GEML thanks NERC for funding. EOB also thanks St. John's College, Oxford, for additional financial support.

\clearpage

\providecommand*\mcitethebibliography{\thebibliography}
\csname @ifundefined\endcsname{endmcitethebibliography}
  {\let\endmcitethebibliography\endthebibliography}{}

\clearpage

\section*{Figures}

\clearpage

\clearpage

\clearpage

\clearpage

\clearpage

\clearpage

\clearpage

\clearpage

\clearpage

\section*{Tables}

\clearpage

\begin{table}
\begin{center}
\begin{tabular}{l l l}
\hline
Parameter & Description & Units\\
\hline
$c_\mathrm{i}$ & Concentration of species i & mol m$^{-3}$\\
\\
$c_\mathrm{i}^{*}$ & Bulk solution concentration of species i & mol m$^{-3}$\\
\\
$d$ & Inter-electrode distance & m\\
\\
$D_\mathrm{i}$ & Diffusion coefficient of species i & mol m$^{-2}$\\
\\
$I$ & Current & A\\
\\
$l$ & Electrode length & m\\
\\
$N$ & Collection efficiency & Unitless\\
\\
$t$ & Time & s\\
\\
$t_N$ & Time taken for collection efficiency to reach $N$ & s\\
\\
$w_e$ & Electrode width & m\\
\\
$x$ & $x$ coordinate & m\\
\\
$y$ & $y$ coordinate & m\\
\\
$z$ & $z$ coordinate & m\\
\hline
\end{tabular}
\end{center}
\caption{Parameter definitions}
\label{DIMENSIONAL}
\end{table}

\clearpage

\begin{table}
\begin{center}
\begin{tabular}{l l}
\hline
Normalised Parameter & Definition\\
\hline
$C_\mathrm{i}$ & $\frac{c_\mathrm{i}}{c^*_\mathrm{A}}$\\
\\
$d'$ & $\frac{d}{w_e}$\\
\\
$D^{'}_\mathrm{i}$ & $\frac{D_\mathrm{i}}{D_\mathrm{A}}$\\
\\
$\tau$ & $\frac{D_\mathrm{A}}{w_e^2}t$\\
\\
$\tau_N$ & $\frac{D_\mathrm{A}}{w_e^2}t_N$\\
\\
$X$ & $\frac{x}{w_e}$\\
\\
$Y$ & $\frac{y}{w_e}$\\
\hline
\end{tabular}
\end{center}
\caption{Normalised parameter definitions}
\label{DIMENSIONLESS}
\end{table}

\end{document}